\documentclass[fleqn,twoside]{article}
\usepackage{espcrc2}

\newcommand{\AmS}{{\protect\the\textfont2
  A\kern-.1667em\lower.5ex\hbox{M}\kern-.125emS}}

\title{Searching for Lorentz Violation}
       
\author{Roland E. Allen and Seiichirou Yokoo \\
Physics Department, Texas A\&M University, College Station, 
Texas 77843}

\begin{document}

\begin{abstract}
Astrophysical, terrestrial, and space-based searches for Lorentz violation 
are very briefly reviewed. Such searches are motivated by the fact that all 
superunified theories (and other theories that attempt to include quantum 
gravity) have some potential for observable violations of Lorentz invariance. 
Another motivation is the exquisite sensitivity of certain well-designed 
experiments and observations to particular forms of Lorentz violation. We also 
review some new predictions of a specific Lorentz-violating theory: If a 
fundamental energy $\overline{m}c^{2}$ in this theory lies below the usual 
GZK cutoff $E_{GZK}$, the cutoff is shifted to infinite energy; i.e., it 
no longer exists. On the other hand, if $\overline{m}c^{2}$ lies above 
$E_{GZK}$, there is a high-energy branch of the fermion dispersion relation 
which provides an alternative mechanism for super-GZK cosmic-ray protons.
\vspace{1pc}
\end{abstract}

\maketitle

\section{Introduction}

During the past few years there has been increasingly widespread
interest in possible violations of Lorentz invariance~[1-29]. There are
several motivations for this interest.

{\bf Theoretical}: Every current candidate for a superunified theory 
contains some potential for Lorentz violation, and the same is 
true for more restricted theories which attempt to treat quantum 
gravity alone. (By a ``superunified theory'' we mean one which includes 
all known physical phenomena, and which is valid up to the Planck energy.) 
Theories with potential for Lorentz violation include superstring/M/brane 
theories, canonical and loop quantum gravity, noncommutative spacetime 
geometry, nontrivial spacetime topology, discrete spacetime structure at 
the Planck length, a variable speed of light or variable physical 
constants, various other \textit{ad hoc} theories, including one that 
specifically addresses the GZK cutoff~\cite{coleman-glashow}, 
and a fundamental theory which will be considered later in this 
paper~\cite{allen1997}. Even in a theory which has Lorentz 
invariance at the most fundamental level, this symmetry can be spontaneously
broken if some field acquires a vacuum expectation value which breaks
rotational invariance or invariance under a boost. (It should
be mentioned that cosmology already provides a preferred frame of reference
-- namely a comoving frame, in which the cosmic background radiation
does not have a dipole anisotropy -- but this is not considered to be 
a breaking of Lorentz symmetry.) A second mechanism for
Lorentz violation is the ``quantum foam'' of Hawking and Wheeler, originally 
envisioned in the context of canonical or path-integral quantization of 
Einstein gravity, but now generalized to other theories with quantum 
gravity. A third possibility is a theory in which Lorentz invariance is not 
postulated to be an exact fundamental symmetry, but instead emerges 
as a low-energy symmetry~\cite{allen1997}.

{\bf Experimental}: Both terrestrial~[3-14] and space-based~[15-20] 
experiments have been designed with exquisite precision which would 
permit detection of even tiny deviations from certain aspects of Lorentz 
invariance. The systems include atoms, charged particles in traps, masers, 
cavity-stabilized oscillators, muons, neutrons, kaons, and other 
neutral mesons.

{\bf Observational}: Particles traveling over cosmological distances from
bright sources (including pulsars, supernovae, blazars, and gamma ray
bursters) allow long-baseline tests which are again sensitive to
even tiny deviations from particular forms of Lorentz 
violation~[21-26]. 

Recall that Lorentz invariance in the context of general relativity means 
\textit{local} Lorentz invariance, or an invariance of the action under 
rotations and boosts involving locally inertial frames of reference. There is 
clearly a connection with the equivalence principle, which can
also be tested in, e.g., space-based experiments. There is a close 
connection with CPT invariance as well: According to the CPT theorem, Lorentz
invariance implies CPT invariance (with the supplementary assumptions of
unitarity and locality). It follows that CPT violation implies Lorentz
violation, although the reverse is not necessarily true. Finally, there is a 
connection to the spin-statistics theorem, which follows from Lorentz 
invariance and microcausality.

We know that P (in the 1950Õs) and CP (in the 1960Õs) have previously 
been found not to be inviolate symmetries, for reasons that are now understood 
in terms of the standard electroweak theory and the CKM matrix. Perhaps 
CPT and Lorentz symmetry are also not inviolate. 

The most extensive theoretical program for systematizing potential 
forms of Lorentz violation and their experimental signals is that of 
Kosteleck\'{y} and coworkers~[3,4,9-18,20,26]. Their philosophy is to 
add small phenomenological Lorentz-violating terms to the Lagrangian of 
the Standard Model, and then interact with a wide variety of experiments 
that can detect such deviations from exact Lorentz or CPT invariance. 
The point of view of this group is rather conservative: The fundamental 
theory (e.g., string theory) is pictured as Lorentz-invariant, with 
Lorentz or CPT violation arising from some form of symmetry-breaking -- 
for example, with a vector field or more general tensor field acquiring a 
vacuum expectation value. Their work has stimulated a considerable amount of 
experimental activity, with further experiments planned for both 
terrestrial and space-based laboratories.

So far there is no undisputed evidence for Lorentz violation, and 
the only solid results from both experiment and observation are strong 
constraints on particular ways in which this symmetry might be broken. 
As an example of an astrophysical constaint, we mention a recent 
paper by Stecker and Glashow~\cite{Stecker1}, in which they conclude 
``ÒWe use the recent reanalysis of multi-TeV [up to ~20 TeV] gamma-ray 
observations of [the blazar] Mrk 501 to constrain the Lorentz invariance
breaking parameter involving the maximum electron velocity. Our limit is 
two orders of magnitude better than that obtained from the maximum observed 
cosmic-ray electron energy.'' Their analysis involves the processes
\begin{equation}
\gamma + \gamma_{infrared} \rightarrow e^{+} +e^{-} \quad \mbox{if} 
\quad c_{e} >  c_{\gamma}
\end{equation}
which can lead to inconsistency with the observation of 20 TeV photons and
\begin{equation}
\gamma \rightarrow e^{+} +e^{-} \quad \mbox{if} 
\quad c_{e} <  c_{\gamma}
\end{equation}
which can lead to inconsistency with the observation of 50 TeV photons.

Another example of astrophysical constraints is the series of analyses by 
Jacobson et al.~[21-24]. In Ref. 22, Jacobson, Liberati, Mattingly, and 
Stecker state ``We strengthen the constraints on possible Lorentz 
symmetry violation (LV) of order $E/M_{Planck}$ for electrons and photons 
in the framework of effective field theory (EFT). The new constraints use 
(i) the absence of vacuum birefringence in the recently observed polarization 
of MeV emission from a gamma ray burst and (ii) the absence of vacuum 
\v{C}erenkov radiation from the synchrotron electrons in the Crab nebula, 
improving the previous bounds by eleven and four orders of magnitude 
respectively.''

Jacobson, Liberati, and Mattingly~\cite{Jacobson1} have obtained a very
strong constraint on a dispersion relation with a cubic term 
in the expression for $E^{2}$:
\begin{equation}
E^{2}=p^{2}+p^{3}/M.
\end{equation}
However, the constraint is less stringent for what may be the more 
natural form with a quartic term:
\begin{equation}
E^{2}=p^{2}+p^{4}/M^{2}.
\end{equation} 

Below we will derive the dispersion relation for a fundamental 
Lorentz-violating theory~\cite{allen1997,allen2002,allen2003} 
and will find that it is easily consistent with these constraints, 
since it has a form quite different from either of those above.

Coleman and Glashow~\cite{coleman-glashow} proposed that the limiting 
velocity of protons, electrons, etc. may be very slightly different 
from the speed of light. (See also Ref. 24.) This is an {\it ad hoc} proposal, 
motivated by the apparent absence of a Greisen-Zatsepin-Kuz'min (GZK) 
cutoff: Ultrahigh energy cosmic ray protons colliding with the cosmic 
microwave background radiation should produce pions, 
\begin{equation}
p + \gamma_{cmb} \rightarrow p +\pi^{0} .
\end{equation}
There should 
consequently be a cutoff in the spectrum of observed protons at about 
50 EeV (or $5 \times 10^{7}$ TeV), if they were created in processes at 
distances of more than about 100 Mpc. But up to 300 EeV cosmic rays 
(presumably protons) appear to be observed, although this is not entirely 
certain~\cite{Bahcall}, and there are also theoretical ideas for a closer 
origin~\cite{Weiler}.

We conclude by mentioning some reviews of terrestrial and space-based 
experiments.

Two reviews  of atomic experiments to test both Lorentz and CPT symmetries, by 
Bluhm~\cite{bluhm-rev}, describe the following: (1)  Penning trap 
experiments with electrons and positrons, and with protons and antiprotons, 
which look for differences in frequencies or sidereal time variations; 
(2) clock comparison experiments, with ÒclockÓ frequencies typically those of 
hyperfine or Zeeman transitions; (3) hydrogen and antihydrogen experiments 
involving ground-state Zeeman hyperfine transitions (at Harvard) or 
1S-2S transitions (proposed at CERN); (4) a spin-polarized torsion pendulum 
experiment (at the University of Washington); (5) muon and muonium 
experiments.

Two reviews by Russell~\cite{russell-space-clocks} discuss clock-based 
experiments to test Lorentz and CPT invariance in space. Such experiments 
will probe the effects of variations in both orientation and velocity.
Among the systems are H masers, laser-cooled Cs and Rb clocks, and 
superconducting microwave cavity oscillators. A number of specific space 
missions have been planned or proposed.
 
Finally, a review by Kosteleck\'{y}~\cite{kos-rev} contains a discussion 
of experiments involving neutral meson (e.g. kaon) oscillations,
a dual nuclear Zeeman He-Xe maser, and cosmological birefringence,  
in addition to the systems mentioned above.

Now let us turn to a specific Lorentz-violating theory~\cite{allen1997} and 
some of its new predictions~\cite{allen2004}. We begin with the action 
for a single initially massless Weyl fermion field~\cite{allen2002}, 
and with the coupling to gauge fields and variations in $e_{\mu }^{\alpha }$ 
neglected: 
\begin{eqnarray}
S_{1} &=&\int d^{4}x\,{\cal {L}}_{1} \\
{\cal {L}}_{1} &=&\,\frac{1}{2}e\,\psi _{1}^{\dagger }\left( \frac{1}{2M}
\eta ^{\mu \nu }\partial _{\nu }\partial _{\mu }+ie_{\alpha }^{\mu }\sigma
^{\alpha }\partial _{\mu }\right) \psi _{1} \nonumber +h.c.
\end{eqnarray}
Here $M$ is a fundamental mass which is comparable to the Planck mass, $\eta
^{\mu \nu }=diag(-1,1,1,1)$ is the Minkowski metric tensor, $\sigma ^{k}$ is
a Pauli matrix, and $\sigma ^{0}$ is the $2\times 2$ identity matrix. Also, 
$e_{\alpha }^{\mu }$ is the gravitational vierbein, which determines the
gravitational metric tensor $g_{\mu \nu }$ through the relations 
\begin{equation}
g_{\mu \nu }=\eta _{\alpha \beta }e_{\mu }^{\alpha }e_{\nu }^{\beta }\quad
,\quad e_{\alpha }^{\mu }e_{\nu }^{\alpha }=\delta _{\nu }^{\mu }.
\end{equation}
A factor of $e^{-1/2}$ has been absorbed in $\psi _{1}$, where 
\begin{equation}
e=\det \,e_{\mu }^{\alpha }=\left( -\det \,g_{\mu \nu }\right) ^{1/2}.
\end{equation}
Fundamental units are used, with $\hbar=c=1$. 
Finally, ``$h.c."$ means ``Hermitian conjugate'', and ${\cal {L}}_{1}$ has
been written in its more fundamental and manifestly Hermitian form. 
The action (6) is invariant under a rotation, but it is not 
invariant under a Lorentz boost because of the first term. (Recall 
that the transformation matrix $\Lambda_{1/2}$ is unitary for a rotation 
and not for a boost~\cite{peskin}.) At low energies, however, this term 
is negligible and full Lorentz invariance is regained.

As before, we choose the directions of the spacetime coordinate axes
to be such that all the $e_{\alpha }^{\mu }$ are positive. If the term
involving $M$ is neglected, ${\cal {L}}_{1}$ has the form appropriate for a
right-handed field. I.e., in order for $S_{1}$ to be invariant under local
Lorentz transformations at low energy, all the fundamental fermionic fields 
$\psi _{1}$ must be taken to transform as right-handed spinors. This is the
reverse of the usual convention in grand-unified theories, where they are
all taken to be left-handed. However, we can convert $\psi _{1}$ to a
left-handed field through the following well-known procedure~\cite
{peskin,ramond,ross}, which is based on the fact that $\left( \sigma ^{2}\right)
^{2}=1$, $\left( \sigma ^{2}\right) ^{\dagger }=\sigma ^{2}$, $\left( \sigma
^{2}\right) ^{*}=-\sigma ^{2}$, and 
\begin{equation}
\sigma ^{2}\sigma ^{k}\sigma ^{2}=-\left( \sigma ^{k}\right) ^{*}.
\end{equation}
Let 
\begin{equation}
\psi _{L}=\sigma ^{2}\psi _{1}^{*}\quad \mbox{or}\quad \psi _{1}=\left(
\sigma ^{2}\psi _{L}\right) ^{*}\quad
\end{equation}
and substitute into (6), using (in the fourth step below) the fact that 
Grassmann fields like $\psi_{L} $ anticommute: 
\begin{eqnarray}
\hspace{-5cm} &&{\cal {L}}_{1} \nonumber \\
\hspace{-5cm} &=&\frac{1}{2}e\,\left[ \left( \sigma ^{2}\psi _{L}\right)
^{*}\right] ^{\dagger }  \nonumber \\
\hspace{-5cm} && \times \left( \frac{1}{2M}\eta ^{\mu \nu }\partial _{\nu
}\partial _{\mu }+ie_{\alpha }^{\mu }\sigma ^{\alpha }\partial _{\mu
}\right) \left( \sigma ^{2}\psi _{L}\right) ^{*} +h.c.  \nonumber \\
\hspace{-5cm} &=&\frac{1}{2}e\,\left[ \left( \frac{1}{2M}\eta ^{\mu \nu }
\partial _{\nu
}\partial _{\mu }+ie_{\alpha }^{\mu }\sigma ^{\alpha }\partial _{\mu
}\right) \left( \sigma ^{2}\psi _{L}\right) ^{*}\right] ^{\dagger }  
\nonumber \\
\hspace{-5cm} && \times \left( \sigma ^{2}\psi _{L}\right) ^{*} +h.c.  
\nonumber \\
\hspace{-5cm} &=&\frac{1}{2}e\,\left[ \left( \frac{1}{2M}\eta ^{\mu \nu }
\partial _{\nu
}\partial _{\mu }+ie_{\alpha }^{\mu }\sigma ^{\alpha }\partial _{\mu
}\right) ^{*}\left( \sigma ^{2}\psi _{L}\right) \right] ^{T}  \nonumber \\
\hspace{-5cm} && \times \left( \sigma ^{2}\psi _{L}\right) ^{*} +h.c.  
\nonumber \\
\hspace{-5cm} &=&-\frac{1}{2}e\,\left[ \left( \sigma ^{2}\psi _{L}\right) ^{*}
\right] ^{T}  \nonumber \\
\hspace{-5cm} && \times \left[ \left( \frac{1}{2M}\eta ^{\mu \nu }
\partial _{\nu }\partial _{\mu
}-ie_{\alpha }^{\mu }\left( \sigma ^{\alpha }\right) ^{*}\partial _{\mu
}\right) \left( \sigma ^{2}\psi _{L}\right) \right]  \nonumber \\
&& \hspace{5cm} +h.c.  \nonumber \\
\hspace{-5cm} &=&\frac{1}{2}e\,\psi _{L}^{\dagger }
\left( \sigma ^{2}\right) ^{\dagger}  \nonumber \\
\hspace{-5cm} && \times \left[ \left( -\frac{1}{2M}\eta ^{\mu \nu }
\partial _{\nu }\partial _{\mu
}+ie_{\alpha }^{\mu }\left( \sigma ^{\alpha }\right) ^{*}\partial _{\mu
}\right) \left( \sigma ^{2}\psi _{L}\right) \right]  \nonumber \\
&& \hspace{5cm} +h.c.  \nonumber \\
\hspace{-5cm} &=&\frac{1}{2}e\,\psi _{L}^{\dagger }  \nonumber \\ 
\hspace{-5cm} && \times \left[ \left( -\frac{1}{2M}\eta ^{\mu
\nu }\partial _{\nu }\partial _{\mu }+ie_{\alpha }^{\mu }\left( \sigma
^{2}\sigma ^{\alpha }\sigma ^{2}\right) ^{*}\partial _{\mu }\right) \psi
_{L}\right]  \nonumber \\
&& \hspace{5cm} +h.c.  \nonumber \\
\hspace{-5cm} &=&\frac{1}{2}e\,\psi _{L}^{\dagger }  \nonumber \\
\hspace{-5cm} && \times \left[ \left( -\frac{1}{2M}\eta ^{\mu
\nu }\partial _{\nu }\partial _{\mu }+ie_{\alpha }^{\mu }\overline{\sigma }
^{\alpha }\partial _{\mu }\right) \psi _{L}\right] +h.c. \nonumber \\
\end{eqnarray}
where $\overline{\sigma }^{0}=$ $\sigma ^{0}$ and $\overline{\sigma }
^{k}=-\sigma ^{k}$. Then $\psi _{L}$ has the Lagrangian appropriate for a
left-handed field (when the term containing $M$ is neglected), and the
definition (10) implies that it transforms as a left-handed field if $\psi
_{1}$ is required to transform as a right-handed 
field~\cite{peskin,ramond,ross}. 

If $\psi _{L}$ corresponds to a particle with a Dirac mass $m$, it
is coupled through this mass to a right-handed field $\psi _{R}$. (The
origin of this mass -- i.e., the coupling to a Higgs field which acquires a
vev -- is not considered in the present paper.) The Lagrangian density for
this pair of fields is then given by 
\begin{eqnarray}
e^{-1}{\cal L} &=&\psi _{R}^{\dagger }\left( \frac{1}{2M}\eta ^{\mu \nu
}\partial _{\nu }\partial _{\mu }+ie_{\alpha }^{\mu }\sigma ^{\alpha
}\partial _{\mu }\right) \psi _{R} \nonumber \\
&&+\psi _{L}^{\dagger }\left( -\frac{1}{2M}\eta ^{\mu \nu }\partial _{\nu
}\partial _{\mu }+ie_{\alpha }^{\mu }\overline{\sigma }^{\alpha }\partial
_{\mu }\right) \psi _{L} \nonumber \\
&&\hspace{1.5cm} -m\psi _{R}^{\dagger }\psi _{L}-m\psi _{L}^{\dagger }\psi _{R}
\end{eqnarray}
after an integration by parts to get the more familiar form. The resulting
equations of motion can be written as 
\begin{eqnarray*}
\left[ \frac{1}{2M}\left( -\,e_{\alpha }^{0}e_{0}^{\alpha }\partial
_{0}\partial _{0}+e_{\alpha }^{k}e_{l}^{\alpha }\partial _{l}\partial
_{k}\right) +ie_{\alpha }^{\mu }\sigma ^{\alpha }\partial _{\mu }\right]
\psi _{R} \\
-m\psi _{L} = 0  \\
\left[ -\frac{1}{2M}\left( -\,e_{\alpha }^{0}e_{0}^{\alpha }\partial
_{0}\partial _{0}+e_{\alpha }^{k}e_{l}^{\alpha }\partial _{l}\partial
_{k}\right) +ie_{\alpha }^{\mu }\overline{\sigma }^{\alpha }\partial _{\mu
}\right] \psi _{L} \\
-m\psi _{R} = 0  
\end{eqnarray*}
with $k,l=1,2,3$. For simplicity, let us assume spatial isotropy and write 
\begin{equation}
e_{\alpha }^{k}=\lambda \delta _{\alpha }^{k}\quad ,\quad e_{k}^{\alpha
}=\lambda ^{-1}\delta _{k}^{\alpha }=\lambda ^{-2}e_{\alpha }^{k}
\end{equation}
\begin{equation}
e_{\alpha }^{0}=\lambda _{0}\delta _{\alpha }^{0}\quad ,\quad e_{0}^{\alpha
}=\lambda _{0}^{-1}\delta _{0}^{\alpha }=\lambda _{0}^{-2}e_{\alpha }^{0}.
\end{equation}
After transforming to a locally inertial frame of reference, in which 
$e_{\alpha }^{\mu }=\delta _{\alpha }^{\mu }$, we have 
\begin{eqnarray}
\left[ \left( -\,\beta \partial _{0}\partial _{0}+\alpha \partial
_{k}\partial _{k}\right) +i\left( \sigma ^{0}\partial _{0}+\sigma
^{k}\partial _{k}\right) \right] \psi _{R} \nonumber \\
-m\psi _{L} = 0 \hspace{0.2cm} \\
\left[ -\left( -\beta \,\partial _{0}\partial _{0}+\alpha \partial
_{k}\partial _{k}\right) +i\left( \sigma ^{0}\partial _{0}-\sigma
^{k}\partial _{k}\right) \right] \psi _{L} \nonumber \\
-m\psi _{R} = 0 \hspace{0.2cm} 
\end{eqnarray}
where 
\begin{equation}
\alpha =\left( 2\lambda ^{2}M\right) ^{-1}\quad ,\quad \beta =\left(
2\lambda _{0}^{2}M\right) ^{-1}.
\end{equation}
At fixed energy $E$ and 3-momentum $\overrightarrow{p}$, these become 
 \begin{eqnarray}
\overrightarrow{\sigma }\cdot \overrightarrow{p}\psi _{R} &=&\left[ \left(
\beta E^{2}-\alpha p^{2}\right) +E\right] \psi _{R}-m\psi _{L} \\
\overrightarrow{\sigma }\cdot \overrightarrow{p}\psi _{L} &=&\left[ \left(
\beta E^{2}-\alpha p^{2}\right) -E\right] \psi _{L}+m\psi _{R}
\end{eqnarray}
where $p$ is the magnitude of $\overrightarrow{p}$, or, since 
$\left( \overrightarrow{\sigma }\cdot \overrightarrow{p}\right)^{2}=p^{2}$, 
\begin{eqnarray}
\left [ \left( p^{2}+m^{2}\right) -\left[ \left( \beta E^{2}-\alpha
p^{2}\right) +E\right] ^{2}\right ] \psi _{R} && \nonumber \\
= -2m\left( \beta E^{2}-\alpha p^{2}\right) \psi _{L} && \\
\left [ \left( p^{2}+m^{2}\right) -\left[ \left( \beta E^{2}-\alpha
p^{2}\right) -E\right] ^{2}\right ] \psi _{L} && \nonumber \\
= 2m\left( \beta E^{2}-\alpha p^{2}\right) \psi _{R} . &&
\end{eqnarray}
We then obtain
\begin{eqnarray}
A_{+}\,A_{-}\, &=&-\left[ 2m\left( \beta E^{2}-\alpha p^{2}\right) \right]
^{2} \\
A_{+} &=&\left( p^{2}+m^{2}\right) -\left[ \left( \beta E^{2}-\alpha
p^{2}\right) +E\right] ^{2} \\
A_{-} &=&\left( p^{2}+m^{2}\right) -\left[ \left( \beta E^{2}-\alpha
p^{2}\right) -E\right] ^{2}
\end{eqnarray}
and (discarding the unphysical root)
\begin{eqnarray}
E^{2} &=&\left( p^{2}+m^{2}\right) +\left( \beta E^{2}-\alpha p^{2}\right)  
\nonumber \\
&&\times \left[ 2\left( E^{2}-m^{2}\right) ^{1/2}-\left( \beta E^{2}-\alpha
p^{2}\right) \right] . \hspace{-2cm} 
\end{eqnarray}
If $m^{2}$ is neglected, (22)-(24) imply that the solutions are 
\begin{eqnarray}
E &=&\mp \frac{1}{2\beta }\pm \frac{1}{2\beta }\left[ 1+4\beta \left( \alpha
p^{2}\pm p\right) \right] ^{1/2} \\
&=&\mp \frac{1}{2\beta }\pm \frac{1}{2\beta }\left[ \left( 1\pm 2\beta
p\right) ^{2}+4\beta \gamma p^{2}\right] ^{1/2}
\end{eqnarray}
where $\gamma =\alpha -\beta $ and the signs are independent.

The various solutions lead to interesting possibilities for new physics which
will be considered in detail elsewhere. For the moment, however, 
consider only the normal branch, for which the first sign is $-$ 
and the last two signs are both $+$. The velocity is then
\begin{eqnarray}
v &=&\partial E / \partial p \\
&=&\left[ \left( 1+2\beta p\right) ^{2}+4\beta \gamma p^{2}\right] ^{-1/2} 
\left( 1+2\beta p+2\gamma p\right)  \nonumber \\
&=&\left[ 1+4\gamma \frac{p+\alpha p^{2}}{1+4\beta p +4\beta
\alpha p^{2}}\right] ^{1/2}.
\end{eqnarray}
It follows that 
\begin{equation}
v > 1 \quad \mbox{if}\quad \alpha >\beta \quad \mbox{and} \quad 
v < 1 \quad \mbox{if}\quad \alpha <\beta .
\end{equation}
As we will find  below, the first possibility would imply vacuum 
\v{C}erenkov radiation, and the second pair production in vacuum, 
so the only plausible possibility is 
\begin{equation}
\alpha =\beta \quad \mbox{which implies that}\quad v=1.
\end{equation}
(In the present paper we do not try to explain the origin of this condition,
but simply accept it as a phenomenological constraint on a 
cosmological scale, far from local gravitational sources.) Then (27) becomes 
\begin{eqnarray}
E &=&\frac{\overline{m}}{2}\,\left[ \mp 1\pm \left( 1\pm 
\frac{2}{\overline{m}}p\right) \right] \\
&=&p,-p,-\overline{m}+p,-\overline{m}-p, 
\overline{m}+p,\overline{m}-p,p,-p \nonumber 
\end{eqnarray}
where 
\begin{equation}
\overline{m}=\beta ^{-1}.
\end{equation}
All massless particles thus travel at the speed of light $c=1$. As 
usual, the destruction operators for negative energies are 
reinterpreted as creation operators for antiparticles with positive 
energies~\cite{allen2002}. The implications of negative group velocities 
for particles and antiparticles will be considered elsewhere, and 
the existence of very high-energy branches in the dispersion relation 
will be discussed below.

For a nonzero mass, but with $\beta = \alpha$, (25) gives 
\begin{eqnarray}
E^{2} &=&\left( p^{2}+m^{2}\right) +\alpha \left( E^{2}-p^{2}\right)  
\nonumber \\
&&\times \left[ 2\left( E^{2}-m^{2}\right) ^{1/2}-\alpha \left(
E^{2}-p^{2}\right) \right] . \hspace{-2cm}
\end{eqnarray}
We are primarily interested in particles with large energy, for which $m^{2}$
(or more precisely $m^{2}/p^{2}$) can be treated as a perturbation: 
\begin{equation}
E^{2}=\left[ E^{2}\right] _{m^{2}=0}+\left[ \partial E^{2} / 
\partial m^{2} \right] _{m^{2}=0}m^{2}.
\end{equation}
From (34) we obtain 
\begin{eqnarray}
&&\partial E^{2}/\partial m^{2}=\left[ 1-\alpha \left(
E^{2}-p^{2}\right) |E|^{-1}\right]  \nonumber \\
&&\times \Big [ 1+\alpha \big[ \alpha \left( E^{2}-p^{2}\right) -2|E| 
\nonumber \\
&&+\left( E^{2}-p^{2}\right) \left( \alpha -|E|^{-1}\right) \big] \Big ] 
^{-1}
\end{eqnarray}
when $\partial E^{2}/\partial m^{2}$ is evaluated at $m=0$. For the 
solutions with $E^{2}=p^{2}$ (when $m=0$), this becomes 
\begin{equation}
\left[ \frac{\partial E^{2}}{\partial m^{2}}\right] _{m^{2}=0}=\left[
1-2\alpha p\right] ^{-1}
\end{equation}
or 
\begin{equation}
E^{2}=p^{2}+\frac{m^{2}}{1-2\alpha p}
\end{equation}
to lowest order in $m^{2}/p^{2}$, 
which reproduces the usual result $E^{2}=p^{2}+m^{2}$ as 
$\alpha p\rightarrow 0$. The particle velocity is then 
$v=\partial E/\partial p = \left ( \partial E^{2}/\partial p \right )
/\left ( 2E \right )$. or
\begin{eqnarray}
v &=&\left[ 1+\frac{\alpha m^{2}}{p\left( 1-2\alpha p\right) ^{2}}\right] 
\nonumber \\
&& \times \left[ 1+\frac{m^{2}}{p^{2}\left( 1-2\alpha p\right) }\right]
^{-1/2} \\
&\approx& 1-\frac{m^{2}}{2p^{2}}\frac{1-4\alpha p}{\left( 1-2\alpha p\right)
^{2}} \\
&=&1-\frac{m^{2}}{2p^{2}}\left( 1-\frac{1}{\left( \left( 2\alpha p\right)
^{-1}-1\right) ^{2}}\right) 
\end{eqnarray}
so that 
\begin{equation}
v\rightarrow 1\;\mbox{as}\; p\rightarrow \infty 
\end{equation}
and
\begin{equation}
v<1 \; \mbox{for} \; p<\overline{m}/4.
\end{equation}
Furthermore, it is easy to see that particles with $p>\overline{m}/4$ will
be superluminal by only an extremely small amount except when $p$ lies 
in a narrow range of energies near $p=\overline{m}/2$ (i.e., $\alpha p=1/2$): 
Letting $\alpha p=1/2+\delta$ in (41), we obtain
\begin{equation}
v-1\approx \frac{1}{2}\frac{m^{2}}{\overline{m}^{2}}\frac{1}{\delta ^{2}} .
\end{equation}
For example, if $m$ is $\sim 1$ GeV and $\overline{m}$ were $\sim 10^{10}$
TeV, then $\delta \sim 10^{-4}$ would imply that  $\left( v-1\right) \sim
10^{-18}$, and the deviation falls like $1/\delta ^{2}$. 
However, it should also be emphasized that superluminal velocities 
of any size are not a violation of causality in the present 
theory, because all signals still propagate forward in time in the initial 
(preferred) frame of reference.

For the solutions with $E^{2}=\left( \overline{m}+p\right) ^{2}$ (when $m=0$),
we obtain 
\begin{eqnarray}
&&\frac{\partial E^{2}}{\partial m^{2}}=\left[ 1-\left( 1+2\alpha p\right)
\left( 1+\alpha p\right) ^{-1}\right]  \nonumber \\
&&\times \Big [ 1+\big[ \left( 1+2\alpha p\right) -2\left( 1+\alpha p\right) 
\nonumber\\
&&+\left( 1+2\alpha p\right) \left( 1-\left( 1+\alpha p\right) ^{-1}\right) 
\big] \Big ] ^{-1} \\
&=&\left[ 1-\frac{\left( 1+2\alpha p\right) }{\left( 1+\alpha p\right) }
\right]  \left[ \left( 1+2\alpha p\right) -\frac{\left( 1+2\alpha p\right) }
{\left( 1+\alpha p\right) }\right] ^{-1} \nonumber \\
&=&-\frac{1}{1+2\alpha p }
\end{eqnarray}
since $\overline{m}=$ $\alpha ^{-1}$. We then have
\begin{equation}
E^{2}=\left( \overline{m}+p\right) ^{2}-\frac{1}{1+2\alpha p }
m^{2}
\end{equation}
again to lowest order in $m^{2}/p^{2}$, and
\begin{eqnarray}
v &=&\left[ \left( \overline{m}+p\right) +\frac{\alpha }{\left( 1+2\alpha
p\right) ^{2}}m^{2}\right]  \nonumber \\
&&\times \left[ \left( \overline{m}+p\right) ^{2}-\frac{1}{\left( 1+2\alpha
p\right) }m^{2}\right] ^{-1/2} \\
&\approx &1+\frac{\left( 3+4\alpha p\right) }{2\left( 1+\alpha p\right)
^{2}\left( 1+2\alpha p\right) ^{2}}\frac{m^{2}}{\overline{m}^{2}}
\end{eqnarray}
so
\begin{equation}
v\rightarrow 1\;\mbox{as}\;p\rightarrow \infty 
\end{equation}
and
\begin{equation}
v\rightarrow 1+\frac{3}{2}\frac{m^{2}}{\overline{m}^{2}}\equiv v_{0}\;
\mbox{as}\;p\rightarrow 0.
\end{equation}
These particles are then slightly superluminal. For example, if $m$ 
is $\sim 1$ GeV and $\overline{m}$ were $\sim 10^{10}$ TeV, then 
$v_{0}-1$ would be $\sim 10^{-26}$. Again, however, a superluminal 
velocity of any size in the present theory does not imply a violation 
of causality.

Now let us turn to the GZK cutoff,~[2,33-38] which results
from collision of a charged fermion with a blackbody photon. The incoming
photon has energy $\omega $ and momentum $(-\omega \cos \theta ,-\omega \sin
\theta ,0)$ in units with $\hbar =c=1$. The incoming fermion has mass $m_{a}$, 
energy $E$, and momentum $(p,0,0)$. The outgoing fermion has mass $m_{b}$, 
energy $E+\omega $, and momentum $(p-\omega \cos \theta ,-\omega \sin \theta
,0)$. If $\omega $ is small (as it is for a blackbody photon), it is valid
to use 
\begin{equation}
\Delta E=\frac{\partial E}{\partial p_{x}}\Delta p_{x}+\frac{\partial E}
{\partial p_{y}}\Delta p_{y}+\frac{\partial E}{\partial m^{2}}\Delta m^{2}
\end{equation}
with $\partial E/\partial p_{k}=v \, p_{k}/p$ and  
$v=\partial E/\partial p$,
so that
\begin{equation}
1+v\cos \theta =\frac{\partial E}{\partial m^{2}}\frac{\Delta m^{2}}{\omega} 
\end{equation}
and the threshold is for a head-on collision. Consider the normal 
branch of the dispersion relation, described by (37), (38), and (41). With
$\partial E/\partial m^{2}=\partial E^{2}/\partial m^{2}/\left ( 2E 
\right )$, (53) becomes
\begin{equation}
2\left( 1+v\cos \theta \right) \left( 1-2\alpha p\right) p=
\Delta m^{2}/\omega 
\end{equation}
where $m^{2}$ has been neglected in comparison to $p^{2}$. This quadratic
equation in $p$ has a solution only if 
\begin{equation}
2\left( 1+v\cos \theta \right) >8\alpha \,\Delta m^{2}/\omega 
\end{equation}
or
\begin{equation}
\overline{m} > 8 \left ( \Delta m^{2}/4 \omega \right)
\end{equation}
since again $\alpha ^{-1}=\overline{m}$. 

If $\overline{m}$ is lower than eight times 
the standard GZK cutoff energy, therefore, 
the present theory implies that the GZK cutoff is eliminated. The reason for
this is that the $\left( 1-2p/\overline{m}\right) $ factor in (38) and (54) 
tends to push the cutoff up to higher energies even if $\overline{m}$ is 
large, and completely eliminates it if $\overline{m}$ falls below 
$2\,\Delta m^{2}/\omega $.

Now consider the high-energy branch of (46), (47), and (49), for which 
$E=\overline{m}+p$ when the mass is neglected. According to (47) and (51), 
particles on this branch travel at essentially the speed of light and 
have an enormous energy
\begin{equation}
E=\sqrt{\overline{m}^{2}-m^{2}}\approx \overline{m}
\end{equation}
even if they have lost essentially all their momentum. 
If such a particle collides with another particle, it can undergo a
transition to the lower branch, with the two particles recoiling in opposite
directions to conserve momentum. Either of them can then enter the Earth's
atmosphere with extraordinary energy comparable to $\overline{m}$. 

If $\overline{m}$ is larger than the standard GZK cutoff energy, 
therefore, the present theory provides an alternative mechanism for cosmic
ray particles above the GZK cutoff. Namely, a particle on the very
high-energy branch (47) can travel cosmological distances without losing more
energy, once it has fallen to the minimum energy $\overline{m}$ for this
branch, and can then undergo a collision relatively near the Earth which
releases this energy.

Finally, let us return to the standard astrophysical threat to a
Lorentz-violating theory, that it may lead to disagreement with the
observations of high-energy matter particles or photons, including
prediction of new processes in the vacuum which are not observed. An example
is vacuum \v{C}erenkov radiation. Conservation of energy and momentum
implies that 
\begin{eqnarray}
-\omega = \Delta E 
= \frac{\partial E}{\partial p_{x}}\Delta p_{x}+\frac{\partial E}
{\partial p_{y}}\Delta p_{y}=\frac{\partial E}{\partial p}\left(
-\omega \cos \theta \right) \nonumber
\end{eqnarray}
so this process can occur if 
\begin{equation}
v=1/\cos \theta \ge 1.
\end{equation}
If we were to have $\beta <\alpha $, the particle velocity at high momentum
would be greater than the velocity of light, and there would be a radiation
of photons in vacuum which is in conflict with 
observation~\cite{coleman-glashow}. 

Next consider the process $photon\rightarrow e^{+}e^{-}$, which
will occur if 
\begin{equation}
2E\left( p\right) =\omega =2p\cos \theta .
\end{equation}
The normal branch for $E\left( p\right) $ corresponds to the choice of
signs $-$, $+$, $+$ in (27). For 20 or 50 TeV photons, it is 
reasonable to assume $\alpha p,\beta p\ll 1$, and keep only the terms of
first and second order in $\alpha $ and $\beta $. Then (27) gives
$E\left( p\right) \approx p+\gamma p^{2}$. 
When the mass term in $E\left( p\right)^{2}$ is also treated only to 
lowest order in $\alpha $ and $\beta $, it is simply $m^{2}$. (E.g., see 
(38).) For a massive particle, therefore, $E\left( p\right)$ becomes 
\begin{eqnarray}
E\left( p\right)  \approx \left[ \left( p+\gamma p^{2}\right)
^{2}+m^{2}\right] ^{1/2} \approx p+\gamma p^{2}+m^{2}/2p \nonumber
\end{eqnarray}
and the condition for vacuum pair production is
\begin{equation}
1+\gamma p+m^{2}/2p^{2}=\cos \theta .
\end{equation}
For $\gamma <0$ this will have a solution if
\begin{equation}
p^{3}>m^{2}/2|\gamma |.
\end{equation}
Since observations indicate that 20 TeV photons do not decay in 
vacuum, $|\gamma |^{-1}$ must lie above the Planck energy.

If $\beta =\alpha $, or $\gamma =0$, the
unphysical processes considered above do not occur. More broadly, 
since many features of Lorentz invariance are retained in the present 
theory (including rotational invariance and the same velocity 
$c$ for all massless particles) it appears that the theory is consistent with 
experiment and observation. The theory is also fundamental, rather 
than {\it ad hoc}, and it leads to various new predictions. Here we 
have emphasized one feature: the behavior of fermions at extremely 
high energy, and the possible implications for the GZK cutoff.

\end{document}